\documentclass[reprint, amsmath,amssymb, aps,super script address]{revtex4-2}
\usepackage{graphicx}
\usepackage{dcolumn}
\usepackage{bm}
\usepackage{amssymb}
\usepackage{amsmath}
\usepackage{appendix}
\usepackage{xcolor}
\usepackage[mathscr]{euscript}
\usepackage{mathtools}
\usepackage{filecontents}
\usepackage{todonotes} 
\usepackage{hyperref}
\usepackage{xcolor}
\usepackage{mathrsfs}
\usepackage{lipsum, babel}
\RequirePackage{filecontents}
\usepackage{cleveref}
\hypersetup{
    colorlinks,
    linkcolor={black!50!black},
    citecolor={black!50!black},
    urlcolor={blue!80!black}
}

\DeclarePairedDelimiter\bra{\langle}{\rvert}
\DeclarePairedDelimiter\ket{\lvert}{\rangle}
\DeclarePairedDelimiterX\braket[2]{\langle}{\rangle}{#1 \delimsize\vert #2}

\begin{document}


\title{Controlled Floquet Dynamics and Topological Bound States in Continuum\\
via Colored Quantum Random Walks}

\author{Zahra Jalali-mola}
\affiliation{School of Physics and CRANN Institute, Trinity College Dublin, University of Dublin, Dublin 2, D02 PN40, Ireland}
\author{Ortwin Hess}
\email{ortwin.hess@tcd.ie}
\affiliation{School of Physics and CRANN Institute, Trinity College Dublin, University of Dublin, Dublin 2, D02 PN40, Ireland}
\affiliation{AMBER, SFI Research Centre for Advanced Materials and BioEngineering Research, Trinity College Dublin, University of Dublin, Dublin 2, D02 PN40, Ireland}

\date{\today}

\begin{abstract}
We demonstrate the emergence and control of Floquet states and topological bound states in the continuum (TBICs) in a two-dimensional colored quantum random walk (cQRW) on a square lattice. By introducing three internal degrees of freedom—termed “colors”—and leveraging SU(3) group representations, we realize dispersive TBICs and intrinsic Floquet dynamics without the need for external periodic driving. Through Chern number calculations, we identify three distinct topological bands, revealing color-induced band mixing as a key mechanism underlying the natural formation of Floquet states. The cQRW framework enables precise tuning of quasi-energy spectra, supporting the emergence of localized edge states in topological band gaps and dispersive TBICs embedded within the bulk of other bands. These TBICs exhibit tunable group velocity, controllable excitation across energy regimes, and robustness, providing theoretical validation for their existence in a first-order Floquet system. Our findings position cQRWs as a powerful platform for investigating and harnessing TBICs and Floquet states, with potential applications in quantum information and communication technologies.
\end{abstract}
\maketitle



Bound states in the continuum (BICs) are localized states that exist within the bulk of electronic or photonic dispersion bands, enabling radiative-loss-free transport and propagation. This unique property has driven advancements in applications such as lasing~\cite{lasig_BIC,Hsu2016,BIC1,Koshelev20}, wavefront manipulation, and nonlinear processes like frequency conversion and mixing~\cite{BSC2023,Kang2023,azzam,Zhigang,Wang2023,Chen2023}.
Expanding on this concept, topological bound states in the continuum (TBICs) harness key features of topological systems—such as bulk-boundary correspondence and edge-state protection—to enable backscatter-free transport and directional propagation~\cite{Yang2013,PhysRevLett.125.213901,PhysRevB.101.161116,Zhigang,sahar,YIN20241660,BIC_NH}. Previous studies have largely focused on TBICs in higher-order or nonlinear topological systems, where their formation is often attributed to symmetry manipulation in the Hamiltonian~\cite{PhysRevLett.125.213901,PhysRevB.101.161116,BIC_NH} or non-Abelian couplings between lattice sites~\cite{Non-Abelian_BIC}. Notably, dispersive TBICs have also been observed in systems comprising two distinct valley topological insulators~\cite{YIN20241660}.

%
\begin{figure}[t]
    \centering
    \includegraphics[width=1\linewidth]{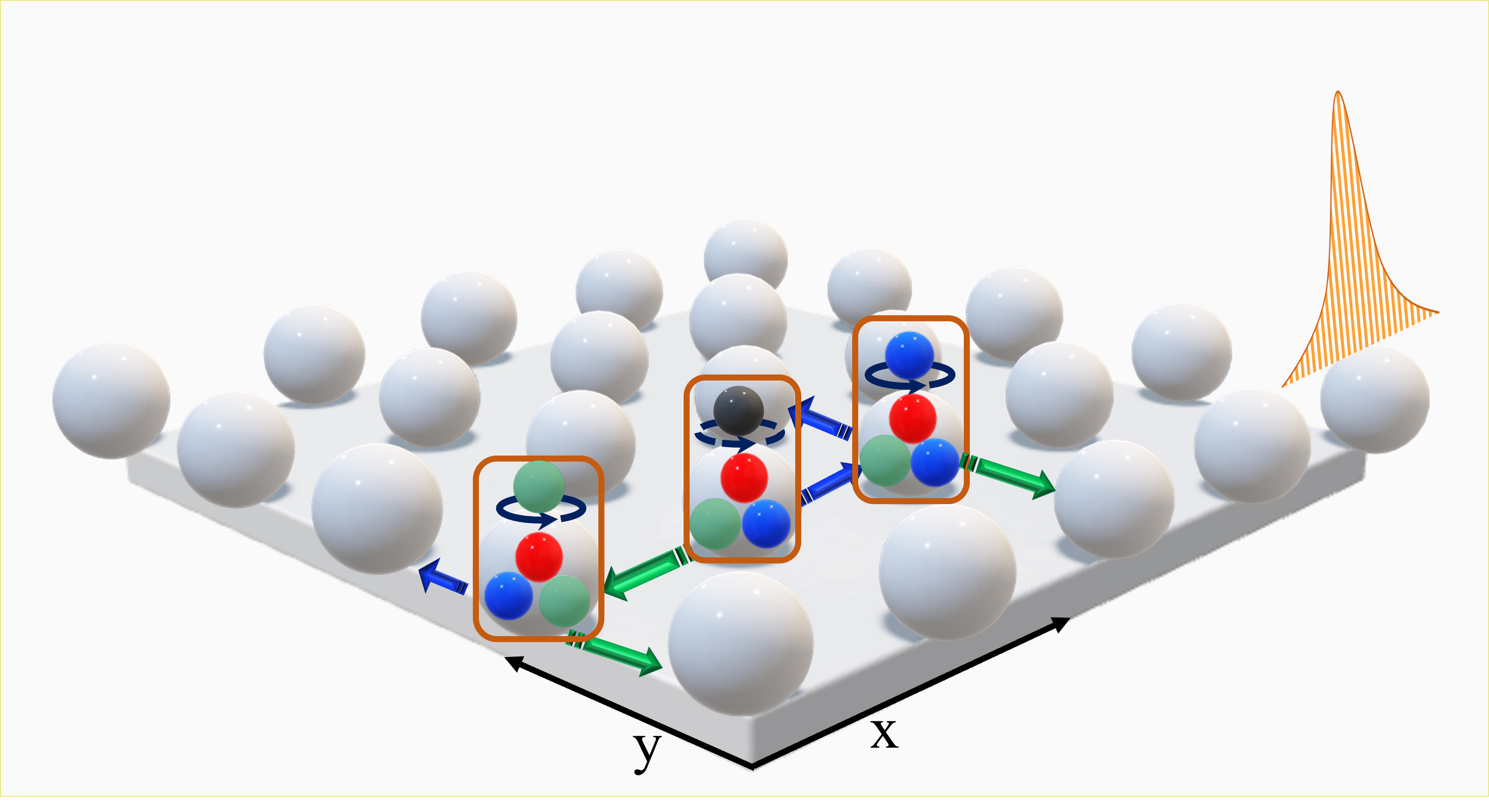}
    \caption{Schematic depiction of a discrete-time cQRW on a square lattice, illustrating the emergence of Floquet dynamics and topological states. The cQRW features three internal states or "colors"—red, blue, and green—associated with movement along the $x(y)$ axis. The walker starts in an initial state, represented by a black sphere, and undergoes a rotation by angle $\theta_1$ through $R(\theta_1)$ (indicated by the dashed circle), generating new color states with varying amplitudes. These amplitudes determine the walker’s movement: blue and green correspond to steps along the $\pm x$ axis, while red corresponds to no movement. In the next step, the new states are rotated by the angle $\theta_2$ via $R(\theta_2)$ (indicated by the solid circle) and move along the $\pm y$ axis according to the same color rule. On the right, a topologically localized edge state, fundamental to the realization of dispersive TBICs and Floquet states, is shown at the boundary.}
    \label{fig:schematic}
\end{figure}
In this work, we demonstrate that lower-order dispersive TBICs naturally emerge in periodically driven Floquet systems realized as colored quantum random walks (cQRWs) on discrete lattice structures. Unlike conventional two-state systems, the cQRW framework introduces three internal degrees of freedom—termed “colors”—which provide unprecedented control over Floquet states and TBICs~\cite{REv_QRW,QRW_aharonov,QRW_Kempe,nayak2000quantumwalkline,Edge-anomalous,Demler_2010,Runder_PRL2009,Runder_PRX2013,Topo_Floquet_PRL2023,Cedzich_2016,PhysRevB.88.121406}. By leveraging the SU(3) group to describe the quantum walker’s spatial evolution, the cQRW offers an intrinsic route to realizing Floquet systems, eliminating the need for external periodic driving.
Through tunable rotation parameters, we uncover unconventional edge-state properties, including localized states within topological band gaps and dispersive TBICs that emerge as edge states embedded within the bulk of other topological bands. These TBICs exhibit tunable group velocity and precise excitation control, arising as intrinsic features of the cQRW’s dynamics.

Conceptually, our approach builds upon previous studies that employ additional degrees of freedom to induce TBIC formation but introduces a novel mechanism: leveraging the randomness and color of the quantum random walk (QRW) to generate TBICs as edge states rather than corner states. This first-order realization of TBICs bridges the gap between higher-order topological states and lower-order systems, providing a controllable platform for investigating the interplay between topology and Floquet dynamics. Moreover, our results establish cQRWs as a versatile framework for robust quantum Floquet systems, opening new avenues for applications in quantum computation and communication technologies~\cite{Quantum_simulation,QRW_Coin,science_computation,Computation_QRW1,Entag_QRW_Ex_2023,QRW_algo1,QRW_algo}.


To realize edge-state TBICs, we consider the motion of a periodically driven quantum random walk (QRW), the quantum mechanical counterpart of a classical random walk. In this system, the dynamics are governed by the internal quantum states of the particle. As in classical random walks, the evolution of the QRW is influenced by a coin operator~\cite{REv_QRW,QRW_aharonov,QRW_Kempe}. However, unlike classical random walks or QRWs with two internal degrees of freedom (e.g., spin or light polarization), we introduce three internal states, referred to as “colors.” These colors, combined with a random shift, define the particle’s motion in a distinctive manner, setting this approach apart from previous models.
The coin operator, modeled as a rotation acting on the internal state, induces a conditional shift, leading to dynamics described by a Floquet Hamiltonian. Under these conditions, the interaction of the QRW with the square lattice enables the formation of TBICs. Crucially, this QRW framework is experimentally feasible in a variety of physical platforms, including trapped ions~\cite{trap_ion_EXP,Blatt2012} and optical lattices~\cite{optical_lattice_QRW}. Furthermore, internal state manipulation can be achieved by illuminating the quantum walker with light carrying different orbital angular momenta, adding further versatility to this approach.


A discrete-time QRW consists of a periodic sequence of unitary operators ($U$) acting on a Hilbert space. Here, $U$ represents a single step of the walk. The Hilbert space of the walker, $\mathcal{H} = \mathcal{H}_p \otimes \mathcal{H}_c$, is composed of two subspaces: $\mathcal{H}_p$, which is spanned by the particle's position states, and $\mathcal{H}_c$, which is the coin space spanned by the quantum states of the coin operator.
In our framework, hopping operations define the particle's movement across a square lattice. The walker is characterized by three mutually orthogonal internal states, termed as red, green, and blue, collectively forming the color basis. The coin operation
is represented by rotations within the SU(3) group space. The walker's dynamics are driven by an alternating sequence of rotation and hopping operations, with adjustable parameters and directions. These dynamics are expressed by the unitary operator:
\begin{align}
    \bm U(\theta_1,\theta_2)=\bm T_y \bm R(\theta_2) \bm T_x \bm R(\theta_1),\quad R(\theta)=e^{i \bm S \theta/2},
\end{align}
  where,  $\bm T_{x/y}$ are the hopping operators that control movement along the $x$ and $y$ directions respectively, and $\bm R(\theta)$ is the coin rotation operator, defined using operators $\bm S$  from the SU(3) symmetry group.

The rotation operator $\bm S$ acts on the internal states of the walker, creating superpositions that influence its translational motion in the $xy$-plane. The translation dynamics, depicted in Fig.\ref{fig:schematic}, incorporate movements to the left and right as well as stationary probabilities and are described as follows\cite{Tude_2022,three-state}:
\begin{equation}
\bm T_i = \ket{i,\mathfrak{R}}\bra{i,\mathfrak{R}} + \ket{i-1,\mathfrak{G}}\bra{i,\mathfrak{G}} + \ket{i+1,\mathfrak{B}}\bra{i,\mathfrak{B}}.
\end{equation}
Here, $\ket{\bm r_i, \text{color}} = \ket{\bm r_i} \otimes \ket{\text{color}}$, where $\bm r_i$ represents position coordinates ($x$ or $y$), and $\text{color}$ denotes the coin or internal basis states. If the walker is in the red state ($\mathfrak{R}$), it remains stationary at position $\bm r$, while the green state ($\mathfrak{G}$) moves it one unit to the left, and the blue state ($\mathfrak{B}$) shifts it one unit to the right along the $x$ or $y$ axis. The internal orthogonal color states are represented as $\ket{\text{color}} = (\mathfrak{R}, \mathfrak{G}, \mathfrak{B})^\text{T}$, where the superscript “T” denotes the transpose. The coin space, spanned by $\ket{\text{color}}$, exploits the symmetry of the SU(3) group.
Rotations within this group are described using the Gell-Mann matrices, $\bm \lambda_i$~\cite{book-lie,Gellman,SU3}. To preserve the structure and dynamics of the QRW, the rotation operator $\bm S$ is chosen to be unitary, satisfying $\bm S=\bm S^\dagger$:
\begin{equation}
   \bm S=\frac{\sqrt{2}}{3}(\bm\lambda_1+\bm\lambda_4+\bm\lambda_6)+\frac{1}{\sqrt{3}}\bm\lambda_8=\frac{1}{3}
  \begin{bmatrix}
    1 & \sqrt{2} & \sqrt{2}  \\
    \sqrt{2} & 1 & \sqrt{2} \\
    \sqrt{2} & \sqrt{2} & -2
  \end{bmatrix}. 
  \label{Eq:SU_Three}
\end{equation}
Taking into account translational invariance along the $x$ and $y$ axes, the shift operators in momentum space are given by:
\begin{equation}
\bm T_l(k) = \text{diag}[1, \exp(i k_l), \exp(-i k_l)], \quad l = x, y,
\end{equation}
where $\bm k=(k_x,k_y)$ represents the momentum vector within the first Brillouin zone. Accordingly, the unitary operator in momentum space is defined as:
\begin{equation}
\label{Uk.eqn}
\bm U_{\bm{k}}(\theta_1,\theta_2)=\bm T(k_y) \bm R(\theta_2) \bm T(k_x) \bm R(\theta_1).
\end{equation}

Following Floquet band theory~\cite{Runder_PRX2013,Runder_PRL2009,Demler_2010}, the effective static Hamiltonian of the quantum walk, denoted as $\bm H(\theta_1,\theta_2)$  can be derived through 
\begin{equation}
    \label{h.eqn}
    e^{-i \bm H(\theta_1,\theta_2)t/\hbar}=\bm U_{\bm{k}}(\theta_1,\theta_2),
\end{equation}
where we adopt units $t/\hbar=1$, corresponding to the time duration of a unitary operator. The quasienergy bands of the bulk Hamiltonian can be obtained from Eq.~\eqref{h.eqn} using the unitary operator defined in Eq.~\eqref{Uk.eqn}. 
Floquet theory predicts that the energy spectrum exhibits periodicity along the energy axis, similar to the periodic structure of momentum space within the first Brillouin zone. As a result, an analogous "energy" Brillouin zone emerges, within the energy range $-\pi<E(k_x,k_y)<\pi$.

By varying the parameters $(\theta_1,\theta_2)$, the system exhibits a wide range of band structures, demonstrating remarkable flexibility and control over the quasi-energy spectrum. This tunability enables the exploration of various phenomena, including band crossings, topological phase transitions, and the emergence of energy gaps within the spectrum. The interplay between these parameters provides valuable insight into the system’s underlying physics and facilitates the design of tailored energy band configurations.
 \begin{figure}[t]
    \centering
    \includegraphics[width=1\linewidth]{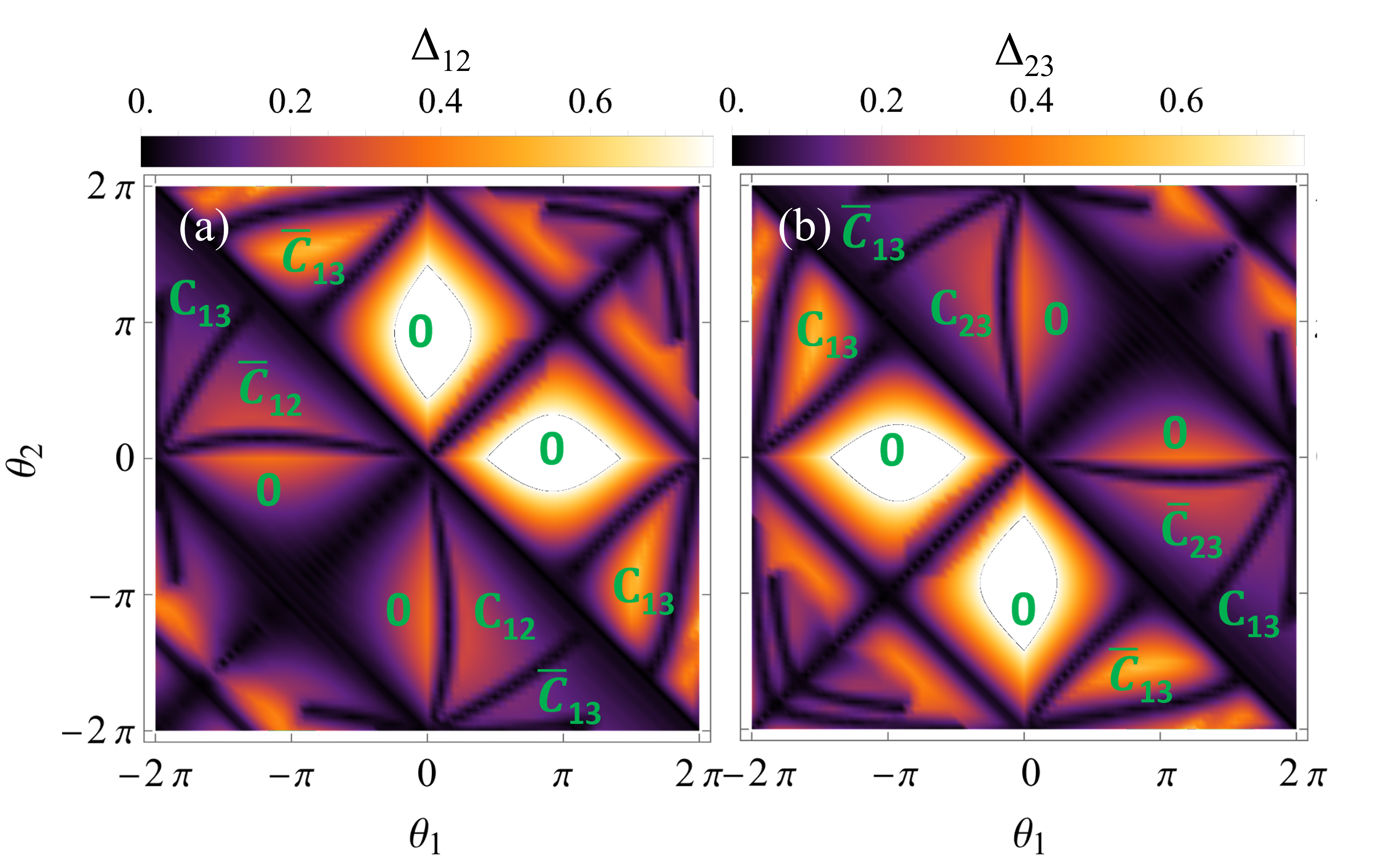}
    \caption{Energy gap and topological number of the bulk Hamiltonian in the rotation plane $(\theta_1,\theta_2)$. Panel~(a) shows the energy gap between the first and second bands, $\Delta_{12}$, while the panel~(b) displays the gap between the second and third bands, $\Delta_{23}$. The color bar indicates the magnitude of the energy gap, in which from  bright to dark energy gaps get smaller and the darker regions signifying gap-closing boundaries. The topological number of each region, represented by $C_{ij}=(C_i=1,C_j=-1)=-\Bar{C}_{ij}$, corresponds to the Chern number of the bands $i, j$. Regions with zero values indicate that the Chern number of at least one band is zero. 
    }
    \label{fig:gap}
\end{figure}

To further investigate the topological properties of the system, we numerically compute the bulk Hamiltonian using Eq.~\eqref{h.eqn}. Our analysis reveals that the eigenenergies do not satisfy the conditions $E_\alpha(\bm{k}) = E_\alpha(-\bm{k})$ or $E_\beta(\bm{k}) = -E_\alpha(-\bm{k})$ $(\alpha \neq \beta)$, confirming the absence of time-reversal and particle-hole symmetries for any nonzero rotation parameters $(\theta_1, \theta_2)$. This symmetry breaking originates from the coin operator, which introduces complex phases via rotation.
Interestingly, the eigenenergies exhibit a residual symmetry, satisfying $E(k_x, k_y) = E(k_y, k_x)$, which reflects an interchange symmetry in momentum space. To further characterize the system’s topology, we compute the Chern number for each energy band by numerically evaluating the band curvature, following the approach in~\cite{Fuki}.

%
\begin{align}
    \label{Berry.eqn}
    & B_n(\bm k)=i\sum_{x,y}\epsilon_{xy}\braket{
    \partial_x \psi_n(\bm k)}{\partial_y \psi_n(\bm k)},\quad \epsilon_{xy}=-\epsilon_{yx}=1 \nonumber\\&
    C_n=-\frac{1}{2\pi}\int_{BZ} d^2\boldsymbol{k} B_n(\bm k), 
\end{align}
 where $\psi_n(\bm k)$ represents the eigenstate of the $n^{\text{th}}$ quasi-energy band of the bulk Hamiltonian. 
 The calculated Chern numbers for each energy band as a function of the rotation parameters $(\theta_1, \theta_2)$ are shown in Fig.~\ref{fig:gap}, marked by green numbers in the center of each gapped region.
 Figure~\ref{fig:gap}(a) illustrates the energy gap $\Delta_{12}$ between the first and second bands between the first and second bands ($\Delta_{ij} = |E_i - E_j|$, while Fig.~\ref{fig:gap}(b) shows the gap $\Delta_{23}$ between the second and third bands. The color gradient indicates the magnitude of the energy gap, with darker shades corresponding to smaller gaps and brighter shades indicating larger gaps. Darker regions, where the gap closes, delineate the boundaries between regions with distinct topological phases.
The Chern number of topological bands $i,j$ within each region is expressed as $C_{ij} = (C_i, C_j) = (1, -1)$, with representative values such as $(1, -1)$ or their opposites $(\Bar{C}_{ij} = -C_{ij})$. 
The regions are classified as topologically trivial when the Chern numbers in both panels are zero. Conversely, if one panel displays a zero Chern number while the other does not, the topological nature of the bands is determined by the non-zero panel. 

\begin{figure}[t]
    \centering
    \includegraphics[width=1.\linewidth]{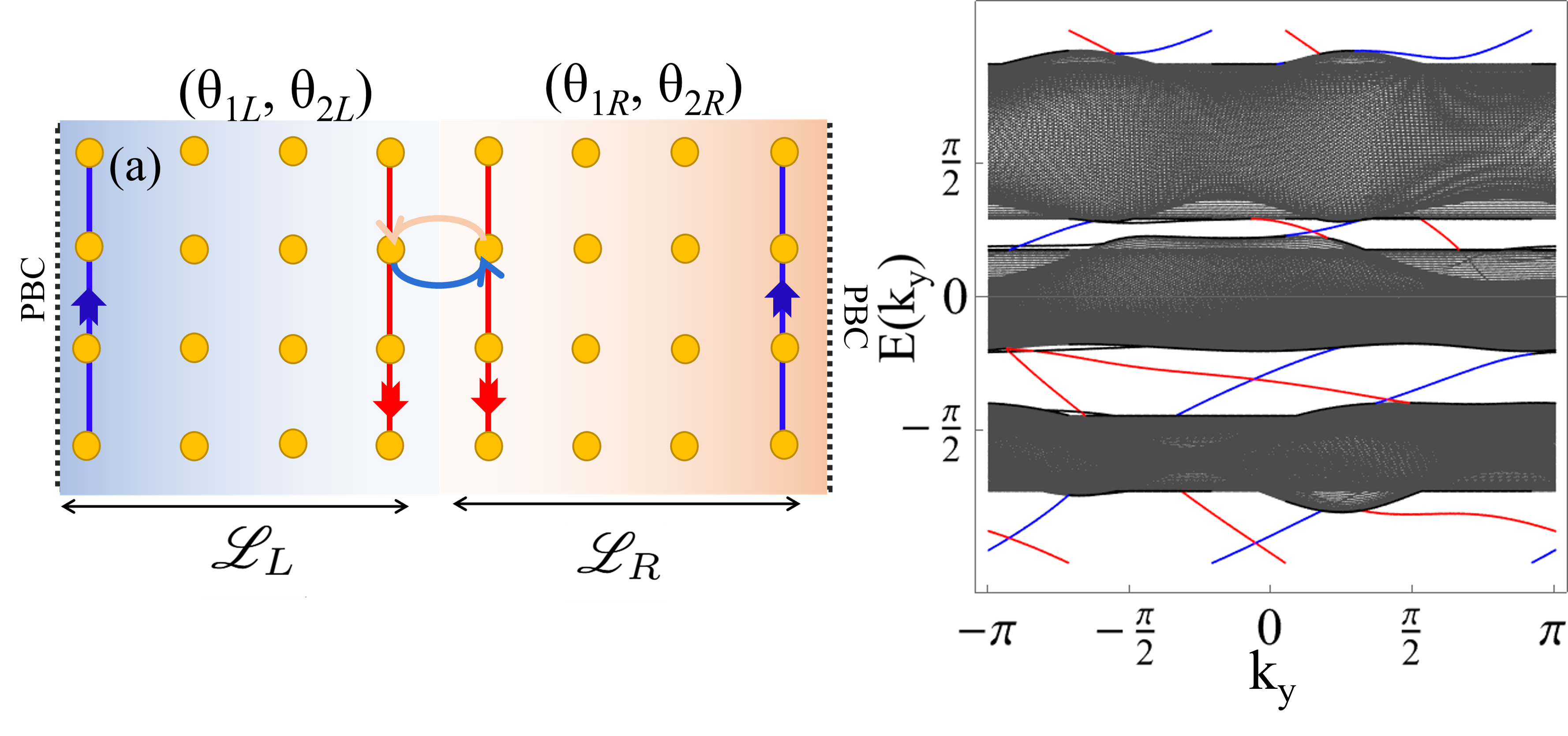}
    \caption{Semi-finite structure of a QRW with finite size along the $x$-axis. Panel~(a) shows the slab configuration, consisting of two distinct media with periodic boundary conditions applied to the external (left and right) boundaries. Panel~(b) presents the energy spectrum of two topologically trivial media
     with zero Chern numbers, highlighting the edge states. Both the left and right media consist of $\mathscr{L}_R=\mathscr{L}_L=60$ lattice sites. The rotation parameters for the left medium are $(\theta_{1L}, \theta_{2L}) = (\pi, \pi/4)$, and for the right medium are $(\theta_{1R}, \theta_{2R}) = (0, \pi)$. Here, the Extended states are shown in black while localized edge states in each band gap are highlighted in blue and red, with their corresponding spatial localization illustrated in panel~(a). The orange and blue arrows in the middle of panel~(a) indicate the transition of rotation parameters across the middle boundary, with blue representing transitions from left to right and orange indicating transitions from right to left, while the rotation parameters at each boundary are determined by their initial position parameters.}
    \label{fig:C=0}
\end{figure}

To further explore the distinct topological phases, we analyze the bulk-boundary correspondence using a semi-finite slab structure.
In this setup, we assume translational invariance along the $y$-axis, while the two media consist of a finite number of lattice sites along the $x$-axis. Periodic boundary conditions are applied to the external boundaries, as depicted in Fig.~\ref{fig:C=0}(a). The topological invariants of the left and right media govern the number of edge states that appear at both the internal and external boundaries of the system.

  We first consider two topologically trivial phases characterized by zero Chern numbers, $C_i=(0,0,0)$ but
  with distinct sets of rotation parameters, as shown in Fig.~\ref{fig:C=0}(b). This configuration includes one interface at the middle and another at the external boundaries due to the periodic boundary conditions.
  At the boundaries, the two media retain their respective rotation parameters at their initial positions. For example, at the inner boundary (indicated by the blue arrow in Fig.~\ref{fig:C=0}(a)), $\theta_{1L}$ governs transitions from the left to the right medium, while $\theta_{1R}$ applies for transitions in the reverse direction. Here, the subscript $L$~($R$) stands for the left~(right) medium. 
  The energy spectrum in Fig.~\ref{fig:C=0}(b), reveals chiral edge states at both the interior and exterior boundaries. In agreement with the bulk-boundary correspondence in Floquet systems~\cite{Runder_PRX2013}, the difference in the number of edge states above and below each band determines the topological invariant associated with that band. In this case, the equal number of edge states around each band indicates that the bands are topologically trivial. The localized edge states at the interior boundary (with negative chirality) and exterior boundary (with positive chirality) are depicted in red and blue, respectively. Importantly, interchanging the left and right media affects only the chirality of these localized edge states.
\begin{figure}[t]
    \centering
    \includegraphics[width=1\linewidth]{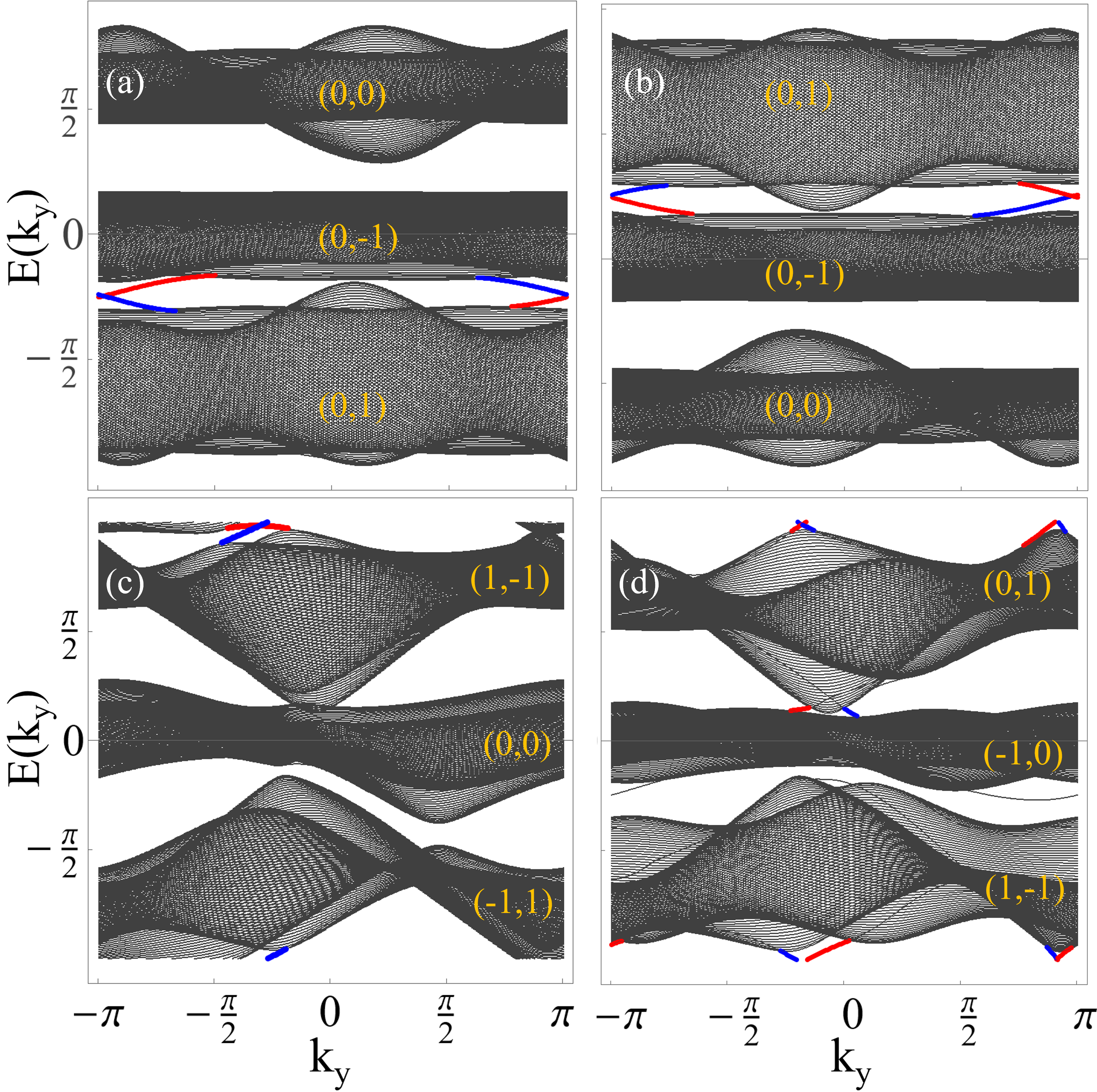}
    \caption{Energy spectrum of a semi-finite structure with two distinct topological phases. The top panels include the left medium with topologically trivial phases, while in the bottom row, both media have different topological invariants. The Chern numbers for each band are displayed within the bandwidth in yellow, as $(C_L,C_R)$ in which subscript $L~(R)$ stands for the left~(right) medium Chern number.  The color coding for localized edge states is the same as Fig.~\ref{fig:C=0}. The rotation parameters for the left and right media, from top left to bottom right, are as follows:  $(\theta_{1L},\theta_{2L},\theta_{1R},\theta_{2R})=$  $(-\pi/32,-\pi,\pi/3,-\pi)$, $(\pi,\pi/32,\pi,-\pi/3)$, 
$(4,-2.5,5.3,-2.5)$, and $(\pi/3,-\pi,4,-2.5)$.}
    \label{fig:edge_T}
\end{figure}


  We now explore scenarios involving media with distinct topological characteristics, ensuring that at least one medium exhibits non-trivial topology. To describe these configurations, we assign Chern numbers to each band as $(C_{L},C_{R})$. 
  Initially, we consider cases where one medium is topologically trivial, while the other possesses non-trivial topology.
  In particular, we examine two distinct non-trivial topological media with Chern numbers  
  $(C_{1R}, C_{2R}, C_{3R}) = (1, -1, 0)$ and $(0, -1, 1)$, each adjacent to a topologically trivial medium. These configurations are depicted in Fig.~\ref{fig:edge_T}(a) and~(b), respectively.
  The presence of localized edge states between the first and second bands in Fig.~\ref{fig:edge_T}(a) implies that these bands have opposing Chern numbers. Conversely, the absence of edge states around the third band's bandwidth indicates its topologically trivial nature.
  In Fig.~\ref{fig:edge_T}(b), however, localized edge states emerge between the second and third bands with reversed chirality compared to panel~(a), reflecting the differing topological numbers. This observation allows us to define the net Chern number for each band as $\tilde{C}_{i} = C_{iR} - C_{iL}$.
  We further investigate configurations involving two non-trivial topological media, as depicted in Fig.~\ref{fig:edge_T}(c) and (d). 
  The quasienergy spectrum depicted in Fig.~\ref{fig:edge_T}(c) shows that the gap between the first and second bands is open, while the gap between the first and third bands is closed. This suggests that, effectively, only one band is topologically trivial. Nonetheless, the topological edge states between the third and first bands remain localized within the bands where the gap has closed. In contrast, Fig.~\ref{fig:edge_T}(d) presents a fully gapped band structure with net Chern numbers calculated as $\tilde{C} = C_{R} - C_{L} = (-2, 1, 1)$. This net Chern number indicates the presence of localized edge states around each band gap. For instance, the lowest energy band has a total Chern number of $\tilde{C}_{1} = -2$, which corresponds to two pairs of localized edge states.
\begin{figure}[t]
    \centering
    \includegraphics[width=1\linewidth]{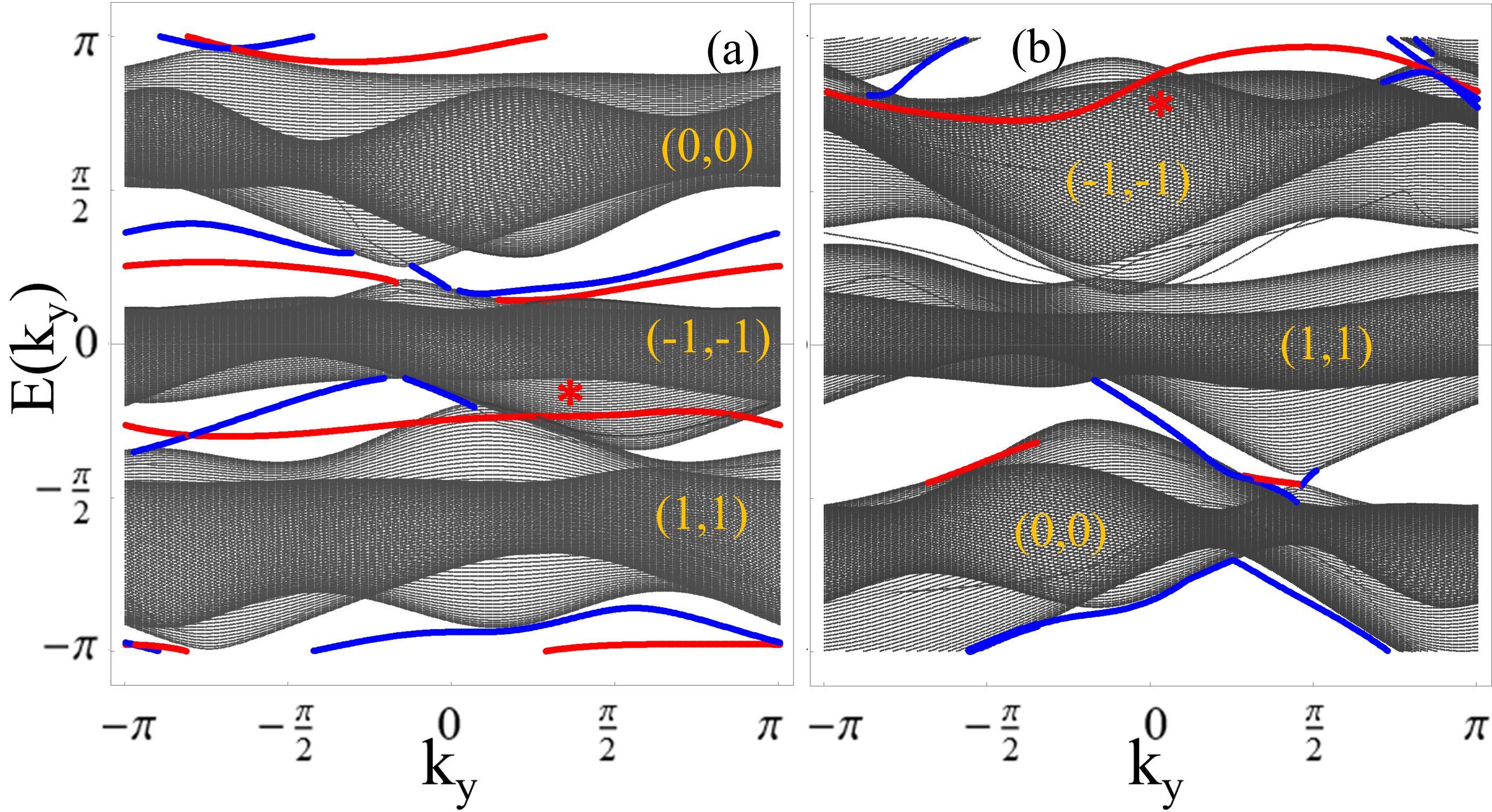}
    \caption{Energy spectrum of semi-infinite structures exhibiting identical non-trivial topological properties but differing rotation parameters in the left and right media. In each panel, the Chern numbers for each band are indicated within the bandwidth in yellow as $(C_L, C_R)$. Localized edge states are color-coded consistently with Fig.~\ref{fig:C=0}. In panel~(a), the rotation parameters are $(\theta_{1L}, \theta_{2L}, \theta_{1R}, \theta_{2R}) = (\pi/3, -\pi, -2\pi, 3\pi)$, corresponding to the topological invariant $(C_L, C_R) = (1, -1, 0)$. In panel~(b), the parameters are $(\theta_{1L}, \theta_{2L}, \theta_{1R}, \theta_{2R}) = (-1.3, \pi, 6, -2.5)$, with $(C_L, C_R) = (0, 1, -1)$. The dispersive TBICs are marked with red stars.}
    \label{fig:edge_STT}
\end{figure}

  After examining various topological configurations of the QRW, we now focus on scenarios where both media share the same topological phase, including potential phase transition points. In Fig.~\ref{fig:edge_STT}(a) and (b), both media exhibit Chern numbers $C = (1, -1, 0)$, $C = (0, 1, -1)$, respectively. 
  The results shown in Fig.~\ref{fig:edge_STT} reveal two notable differences compared to the previous configurations. First, the chiral edge states at the external boundary (blue lines) form pairs with opposite chirality originated from the same topological Chern number of each band at both media, whereas in earlier examples, paired chiral edge states shared the same chirality. Second, since both media possess non-trivial topological characteristics, the localized edge states at the middle boundary lack chiral behavior, and we refer to them as mid-gap states as they do not connect energy bands. These properties make the edge states distinct from that of trivial bands in Fig.~\ref{fig:C=0}(b).  

In Fig.~\ref{fig:edge_STT}, two strongly localized edge states, marked by the red star, coexist with the gapless bands while remaining unhybridized with the bulk extended states. These states appear as dispersive TBICs that are robust against random disorder.
Unlike flat bands and higher-order TBICs, these two TBICs emerge within a semi-infinite configuration and exhibit a dispersive nature. To further investigate the physical characteristics of the BICs, we examine a fully finite system in the $xy$-plane in the Supplemental Material~\cite{supp}, considering a finite configuration of two media, where one medium is enclosed by the other. The results indicate that chiral edge states (depicted in blue) remain localized at the system’s edge, whereas the dispersive TBICs, initially localized at the edge, transform into corner states while still being surrounded by bulk states.

In summary, we have investigated the topological properties of a novel color quantum random walk (cQRW) system on a square lattice, introducing three internal degrees of freedom—termed “colors”—to the quantum walker. This three-state configuration, described by SU(3) group representations, extends beyond conventional two-state models and enables the realization of dispersive topological bound states in the continuum (TBICs). Moreover, Floquet dynamics emerge naturally as an intrinsic property of the system, without requiring external periodic driving.

Our results demonstrate that the cQRW offers a robust and tunable platform for controlling Floquet states. By varying rotation parameters and computing Chern numbers, we identify three distinct topological quasi-energy bands exhibiting phase transitions driven by color-induced mixing. These transitions manifest through bulk-boundary correspondence, which we systematically examine in both semi-infinite and fully finite geometries.

A key outcome of our study is the identification of multiple topological configurations—trivial-trivial, trivial-topological, and topological-topological—each characterized by specific distributions of chiral edge states. The cQRW further enables precise tuning of color-induced band dispersions, revealing diverse states such as chiral edge modes and mid-gap states within topological band gaps. Notably, the TBICs identified here are intrinsically dispersive with tunable group velocity, and their excitation within the quasi-energy spectrum can be precisely controlled across different energy regimes. In finite lattice systems, we further demonstrate that these TBICs transform into bound-in-continuum (BIC) states, manifesting as topological corner states.

This work establishes the theoretical foundation for TBICs in Floquet systems and positions cQRWs as a powerful framework for engineering and manipulating topological states. Our findings open new avenues for leveraging topological quantum walks in quantum technologies, particularly in the design of controllable quantum state transport and storage.

\begin{acknowledgments}
O.H. gratefully acknowledges funding from Research Ireland (formerly Science Foundation Ireland) via the METAQUANT Research Professorship Programme (Grant No. 18/RP/6236). 
\end{acknowledgments}
\bibliography{ref.bib}

\begin{thebibliography}{45}%
\makeatletter
\providecommand \@ifxundefined [1]{%
 \@ifx{#1\undefined}
}%
\providecommand \@ifnum [1]{%
 \ifnum #1\expandafter \@firstoftwo
 \else \expandafter \@secondoftwo
 \fi
}%
\providecommand \@ifx [1]{%
 \ifx #1\expandafter \@firstoftwo
 \else \expandafter \@secondoftwo
 \fi
}%
\providecommand \natexlab [1]{#1}%
\providecommand \enquote  [1]{``#1''}%
\providecommand \bibnamefont  [1]{#1}%
\providecommand \bibfnamefont [1]{#1}%
\providecommand \citenamefont [1]{#1}%
\providecommand \href@noop [0]{\@secondoftwo}%
\providecommand \href [0]{\begingroup \@sanitize@url \@href}%
\providecommand \@href[1]{\@@startlink{#1}\@@href}%
\providecommand \@@href[1]{\endgroup#1\@@endlink}%
\providecommand \@sanitize@url [0]{\catcode `\\12\catcode `\$12\catcode `\&12\catcode `\#12\catcode `\^12\catcode `\_12\catcode `\%12\relax}%
\providecommand \@@startlink[1]{}%
\providecommand \@@endlink[0]{}%
\providecommand \url  [0]{\begingroup\@sanitize@url \@url }%
\providecommand \@url [1]{\endgroup\@href {#1}{\urlprefix }}%
\providecommand \urlprefix  [0]{URL }%
\providecommand \Eprint [0]{\href }%
\providecommand \doibase [0]{https://doi.org/}%
\providecommand \selectlanguage [0]{\@gobble}%
\providecommand \bibinfo  [0]{\@secondoftwo}%
\providecommand \bibfield  [0]{\@secondoftwo}%
\providecommand \translation [1]{[#1]}%
\providecommand \BibitemOpen [0]{}%
\providecommand \bibitemStop [0]{}%
\providecommand \bibitemNoStop [0]{.\EOS\space}%
\providecommand \EOS [0]{\spacefactor3000\relax}%
\providecommand \BibitemShut  [1]{\csname bibitem#1\endcsname}%
\let\auto@bib@innerbib\@empty
\bibitem [{\citenamefont {Song}\ \emph {et~al.}(2020)\citenamefont {Song}, \citenamefont {Hu}, \citenamefont {Dai}, \citenamefont {Zheng}, \citenamefont {Han}, \citenamefont {Zi}, \citenamefont {Zhang},\ and\ \citenamefont {Chan}}]{lasig_BIC}%
  \BibitemOpen
  \bibfield  {author} {\bibinfo {author} {\bibfnamefont {Q.}~\bibnamefont {Song}}, \bibinfo {author} {\bibfnamefont {J.}~\bibnamefont {Hu}}, \bibinfo {author} {\bibfnamefont {S.}~\bibnamefont {Dai}}, \bibinfo {author} {\bibfnamefont {C.}~\bibnamefont {Zheng}}, \bibinfo {author} {\bibfnamefont {D.}~\bibnamefont {Han}}, \bibinfo {author} {\bibfnamefont {J.}~\bibnamefont {Zi}}, \bibinfo {author} {\bibfnamefont {Z.~Q.}\ \bibnamefont {Zhang}},\ and\ \bibinfo {author} {\bibfnamefont {C.~T.}\ \bibnamefont {Chan}},\ }\bibfield  {title} {\bibinfo {title} {Coexistence of a new type of bound state in the continuum and a lasing threshold mode induced by pt symmetry},\ }\href {https://doi.org/10.1126/sciadv.abc1160} {\bibfield  {journal} {\bibinfo  {journal} {Science Advances}\ }\textbf {\bibinfo {volume} {6}},\ \bibinfo {pages} {eabc1160} (\bibinfo {year} {2020})}\BibitemShut {NoStop}%
\bibitem [{\citenamefont {Hsu}\ \emph {et~al.}(2016)\citenamefont {Hsu}, \citenamefont {Zhen}, \citenamefont {Stone}, \citenamefont {Joannopoulos},\ and\ \citenamefont {Solja{\v{c}}i{\'{c}}}}]{Hsu2016}%
  \BibitemOpen
  \bibfield  {author} {\bibinfo {author} {\bibfnamefont {C.~W.}\ \bibnamefont {Hsu}}, \bibinfo {author} {\bibfnamefont {B.}~\bibnamefont {Zhen}}, \bibinfo {author} {\bibfnamefont {A.~D.}\ \bibnamefont {Stone}}, \bibinfo {author} {\bibfnamefont {J.~D.}\ \bibnamefont {Joannopoulos}},\ and\ \bibinfo {author} {\bibfnamefont {M.}~\bibnamefont {Solja{\v{c}}i{\'{c}}}},\ }\bibfield  {title} {\bibinfo {title} {Bound states in the continuum},\ }\href {https://doi.org/10.1038/natrevmats.2016.48} {\bibfield  {journal} {\bibinfo  {journal} {Nature Reviews Materials}\ }\textbf {\bibinfo {volume} {1}},\ \bibinfo {pages} {16048} (\bibinfo {year} {2016})}\BibitemShut {NoStop}%
\bibitem [{\citenamefont {Plotnik}\ \emph {et~al.}(2011)\citenamefont {Plotnik}, \citenamefont {Peleg}, \citenamefont {Dreisow}, \citenamefont {Heinrich}, \citenamefont {Nolte}, \citenamefont {Szameit},\ and\ \citenamefont {Segev}}]{BIC1}%
  \BibitemOpen
  \bibfield  {author} {\bibinfo {author} {\bibfnamefont {Y.}~\bibnamefont {Plotnik}}, \bibinfo {author} {\bibfnamefont {O.}~\bibnamefont {Peleg}}, \bibinfo {author} {\bibfnamefont {F.}~\bibnamefont {Dreisow}}, \bibinfo {author} {\bibfnamefont {M.}~\bibnamefont {Heinrich}}, \bibinfo {author} {\bibfnamefont {S.}~\bibnamefont {Nolte}}, \bibinfo {author} {\bibfnamefont {A.}~\bibnamefont {Szameit}},\ and\ \bibinfo {author} {\bibfnamefont {M.}~\bibnamefont {Segev}},\ }\bibfield  {title} {\bibinfo {title} {Experimental observation of optical bound states in the continuum},\ }\href {https://doi.org/10.1103/PhysRevLett.107.183901} {\bibfield  {journal} {\bibinfo  {journal} {Phys. Rev. Lett.}\ }\textbf {\bibinfo {volume} {107}},\ \bibinfo {pages} {183901} (\bibinfo {year} {2011})}\BibitemShut {NoStop}%
\bibitem [{\citenamefont {Koshelev}\ \emph {et~al.}(2020)\citenamefont {Koshelev}, \citenamefont {Bogdanov},\ and\ \citenamefont {Kivshar}}]{Koshelev20}%
  \BibitemOpen
  \bibfield  {author} {\bibinfo {author} {\bibfnamefont {K.}~\bibnamefont {Koshelev}}, \bibinfo {author} {\bibfnamefont {A.}~\bibnamefont {Bogdanov}},\ and\ \bibinfo {author} {\bibfnamefont {Y.}~\bibnamefont {Kivshar}},\ }\bibfield  {title} {\bibinfo {title} {Engineering with bound states in the continuum},\ }\href {https://doi.org/10.1364/OPN.31.1.000038} {\bibfield  {journal} {\bibinfo  {journal} {Opt. Photon. News}\ }\textbf {\bibinfo {volume} {31}},\ \bibinfo {pages} {38} (\bibinfo {year} {2020})}\BibitemShut {NoStop}%
\bibitem [{\citenamefont {Huang}\ \emph {et~al.}(2024)\citenamefont {Huang}, \citenamefont {Ke}, \citenamefont {Zhong}, \citenamefont {Kivshar},\ and\ \citenamefont {Lee}}]{BSC2023}%
  \BibitemOpen
  \bibfield  {author} {\bibinfo {author} {\bibfnamefont {B.}~\bibnamefont {Huang}}, \bibinfo {author} {\bibfnamefont {Y.}~\bibnamefont {Ke}}, \bibinfo {author} {\bibfnamefont {H.}~\bibnamefont {Zhong}}, \bibinfo {author} {\bibfnamefont {Y.~S.}\ \bibnamefont {Kivshar}},\ and\ \bibinfo {author} {\bibfnamefont {C.}~\bibnamefont {Lee}},\ }\bibfield  {title} {\bibinfo {title} {Interaction-induced multiparticle bound states in the continuum},\ }\href {https://doi.org/10.1103/PhysRevLett.133.140202} {\bibfield  {journal} {\bibinfo  {journal} {Phys. Rev. Lett.}\ }\textbf {\bibinfo {volume} {133}},\ \bibinfo {pages} {140202} (\bibinfo {year} {2024})}\BibitemShut {NoStop}%
\bibitem [{\citenamefont {Kang}\ \emph {et~al.}(2023)\citenamefont {Kang}, \citenamefont {Liu}, \citenamefont {Chan},\ and\ \citenamefont {Xiao}}]{Kang2023}%
  \BibitemOpen
  \bibfield  {author} {\bibinfo {author} {\bibfnamefont {M.}~\bibnamefont {Kang}}, \bibinfo {author} {\bibfnamefont {T.}~\bibnamefont {Liu}}, \bibinfo {author} {\bibfnamefont {C.~T.}\ \bibnamefont {Chan}},\ and\ \bibinfo {author} {\bibfnamefont {M.}~\bibnamefont {Xiao}},\ }\bibfield  {title} {\bibinfo {title} {Applications of bound states in the continuum in photonics},\ }\href {https://doi.org/10.1038/s42254-023-00642-8} {\bibfield  {journal} {\bibinfo  {journal} {Nature Reviews Physics}\ }\textbf {\bibinfo {volume} {5}},\ \bibinfo {pages} {659} (\bibinfo {year} {2023})}\BibitemShut {NoStop}%
\bibitem [{\citenamefont {Azzam}\ and\ \citenamefont {Kildishev}(2021)}]{azzam}%
  \BibitemOpen
  \bibfield  {author} {\bibinfo {author} {\bibfnamefont {S.~I.}\ \bibnamefont {Azzam}}\ and\ \bibinfo {author} {\bibfnamefont {A.~V.}\ \bibnamefont {Kildishev}},\ }\bibfield  {title} {\bibinfo {title} {Photonic bound states in the continuum: From basics to applications},\ }\href {https://doi.org/https://doi.org/10.1002/adom.202001469} {\bibfield  {journal} {\bibinfo  {journal} {Advanced Optical Materials}\ }\textbf {\bibinfo {volume} {9}},\ \bibinfo {pages} {2001469} (\bibinfo {year} {2021})}\BibitemShut {NoStop}%
\bibitem [{\citenamefont {Hu}\ \emph {et~al.}(2021)\citenamefont {Hu}, \citenamefont {Bongiovanni}, \citenamefont {Juki{\'{c}}}, \citenamefont {Jajti{\'{c}}}, \citenamefont {Xia}, \citenamefont {Song}, \citenamefont {Xu}, \citenamefont {Morandotti}, \citenamefont {Buljan},\ and\ \citenamefont {Chen}}]{Zhigang}%
  \BibitemOpen
  \bibfield  {author} {\bibinfo {author} {\bibfnamefont {Z.}~\bibnamefont {Hu}}, \bibinfo {author} {\bibfnamefont {D.}~\bibnamefont {Bongiovanni}}, \bibinfo {author} {\bibfnamefont {D.}~\bibnamefont {Juki{\'{c}}}}, \bibinfo {author} {\bibfnamefont {E.}~\bibnamefont {Jajti{\'{c}}}}, \bibinfo {author} {\bibfnamefont {S.}~\bibnamefont {Xia}}, \bibinfo {author} {\bibfnamefont {D.}~\bibnamefont {Song}}, \bibinfo {author} {\bibfnamefont {J.}~\bibnamefont {Xu}}, \bibinfo {author} {\bibfnamefont {R.}~\bibnamefont {Morandotti}}, \bibinfo {author} {\bibfnamefont {H.}~\bibnamefont {Buljan}},\ and\ \bibinfo {author} {\bibfnamefont {Z.}~\bibnamefont {Chen}},\ }\bibfield  {title} {\bibinfo {title} {Nonlinear control of photonic higher-order topological bound states in the continuum},\ }\href {https://doi.org/10.1038/s41377-021-00607-5} {\bibfield  {journal} {\bibinfo  {journal} {Light: Science {\&} Applications}\ }\textbf {\bibinfo {volume} {10}},\ \bibinfo {pages} {164} (\bibinfo {year} {2021})}\BibitemShut {NoStop}%
\bibitem [{\citenamefont {Wang}\ \emph {et~al.}(2023)\citenamefont {Wang}, \citenamefont {Srivastava}, \citenamefont {Tan}, \citenamefont {Wang},\ and\ \citenamefont {Singh}}]{Wang2023}%
  \BibitemOpen
  \bibfield  {author} {\bibinfo {author} {\bibfnamefont {W.}~\bibnamefont {Wang}}, \bibinfo {author} {\bibfnamefont {Y.~K.}\ \bibnamefont {Srivastava}}, \bibinfo {author} {\bibfnamefont {T.~C.}\ \bibnamefont {Tan}}, \bibinfo {author} {\bibfnamefont {Z.}~\bibnamefont {Wang}},\ and\ \bibinfo {author} {\bibfnamefont {R.}~\bibnamefont {Singh}},\ }\bibfield  {title} {\bibinfo {title} {Brillouin zone folding driven bound states in the continuum},\ }\href {https://doi.org/10.1038/s41467-023-38367-y} {\bibfield  {journal} {\bibinfo  {journal} {Nature Communications}\ }\textbf {\bibinfo {volume} {14}},\ \bibinfo {pages} {2811} (\bibinfo {year} {2023})}\BibitemShut {NoStop}%
\bibitem [{\citenamefont {Chen}\ \emph {et~al.}(2023)\citenamefont {Chen}, \citenamefont {Deng}, \citenamefont {Sha}, \citenamefont {Chen}, \citenamefont {Wang}, \citenamefont {Chen}, \citenamefont {Wu}, \citenamefont {Chu}, \citenamefont {Kivshar}, \citenamefont {Xiao},\ and\ \citenamefont {Qiu}}]{Chen2023}%
  \BibitemOpen
  \bibfield  {author} {\bibinfo {author} {\bibfnamefont {Y.}~\bibnamefont {Chen}}, \bibinfo {author} {\bibfnamefont {H.}~\bibnamefont {Deng}}, \bibinfo {author} {\bibfnamefont {X.}~\bibnamefont {Sha}}, \bibinfo {author} {\bibfnamefont {W.}~\bibnamefont {Chen}}, \bibinfo {author} {\bibfnamefont {R.}~\bibnamefont {Wang}}, \bibinfo {author} {\bibfnamefont {Y.-H.}\ \bibnamefont {Chen}}, \bibinfo {author} {\bibfnamefont {D.}~\bibnamefont {Wu}}, \bibinfo {author} {\bibfnamefont {J.}~\bibnamefont {Chu}}, \bibinfo {author} {\bibfnamefont {Y.~S.}\ \bibnamefont {Kivshar}}, \bibinfo {author} {\bibfnamefont {S.}~\bibnamefont {Xiao}},\ and\ \bibinfo {author} {\bibfnamefont {C.-W.}\ \bibnamefont {Qiu}},\ }\bibfield  {title} {\bibinfo {title} {Observation of intrinsic chiral bound states in the continuum},\ }\href {https://doi.org/10.1038/s41586-022-05467-6} {\bibfield  {journal} {\bibinfo  {journal} {Nature}\ }\textbf {\bibinfo {volume} {613}},\ \bibinfo {pages} {474} (\bibinfo {year} {2023})}\BibitemShut {NoStop}%
\bibitem [{\citenamefont {Yang}\ \emph {et~al.}(2013)\citenamefont {Yang}, \citenamefont {Saeed~Bahramy},\ and\ \citenamefont {Nagaosa}}]{Yang2013}%
  \BibitemOpen
  \bibfield  {author} {\bibinfo {author} {\bibfnamefont {B.-J.}\ \bibnamefont {Yang}}, \bibinfo {author} {\bibfnamefont {M.}~\bibnamefont {Saeed~Bahramy}},\ and\ \bibinfo {author} {\bibfnamefont {N.}~\bibnamefont {Nagaosa}},\ }\bibfield  {title} {\bibinfo {title} {Topological protection of bound states against the hybridization},\ }\href {https://doi.org/10.1038/ncomms2524} {\bibfield  {journal} {\bibinfo  {journal} {Nature Communications}\ }\textbf {\bibinfo {volume} {4}},\ \bibinfo {pages} {1524} (\bibinfo {year} {2013})}\BibitemShut {NoStop}%
\bibitem [{\citenamefont {Cerjan}\ \emph {et~al.}(2020)\citenamefont {Cerjan}, \citenamefont {J\"urgensen}, \citenamefont {Benalcazar}, \citenamefont {Mukherjee},\ and\ \citenamefont {Rechtsman}}]{PhysRevLett.125.213901}%
  \BibitemOpen
  \bibfield  {author} {\bibinfo {author} {\bibfnamefont {A.}~\bibnamefont {Cerjan}}, \bibinfo {author} {\bibfnamefont {M.}~\bibnamefont {J\"urgensen}}, \bibinfo {author} {\bibfnamefont {W.~A.}\ \bibnamefont {Benalcazar}}, \bibinfo {author} {\bibfnamefont {S.}~\bibnamefont {Mukherjee}},\ and\ \bibinfo {author} {\bibfnamefont {M.~C.}\ \bibnamefont {Rechtsman}},\ }\bibfield  {title} {\bibinfo {title} {Observation of a higher-order topological bound state in the continuum},\ }\href {https://doi.org/10.1103/PhysRevLett.125.213901} {\bibfield  {journal} {\bibinfo  {journal} {Phys. Rev. Lett.}\ }\textbf {\bibinfo {volume} {125}},\ \bibinfo {pages} {213901} (\bibinfo {year} {2020})}\BibitemShut {NoStop}%
\bibitem [{\citenamefont {Benalcazar}\ and\ \citenamefont {Cerjan}(2020)}]{PhysRevB.101.161116}%
  \BibitemOpen
  \bibfield  {author} {\bibinfo {author} {\bibfnamefont {W.~A.}\ \bibnamefont {Benalcazar}}\ and\ \bibinfo {author} {\bibfnamefont {A.}~\bibnamefont {Cerjan}},\ }\bibfield  {title} {\bibinfo {title} {Bound states in the continuum of higher-order topological insulators},\ }\href {https://doi.org/10.1103/PhysRevB.101.161116} {\bibfield  {journal} {\bibinfo  {journal} {Phys. Rev. B}\ }\textbf {\bibinfo {volume} {101}},\ \bibinfo {pages} {161116} (\bibinfo {year} {2020})}\BibitemShut {NoStop}%
\bibitem [{\citenamefont {Jalali-mola}\ \emph {et~al.}(2023)\citenamefont {Jalali-mola}, \citenamefont {Grass}, \citenamefont {Kasper}, \citenamefont {Lewenstein},\ and\ \citenamefont {Bhattacharya}}]{sahar}%
  \BibitemOpen
  \bibfield  {author} {\bibinfo {author} {\bibfnamefont {Z.}~\bibnamefont {Jalali-mola}}, \bibinfo {author} {\bibfnamefont {T.}~\bibnamefont {Grass}}, \bibinfo {author} {\bibfnamefont {V.}~\bibnamefont {Kasper}}, \bibinfo {author} {\bibfnamefont {M.}~\bibnamefont {Lewenstein}},\ and\ \bibinfo {author} {\bibfnamefont {U.}~\bibnamefont {Bhattacharya}},\ }\bibfield  {title} {\bibinfo {title} {Topological bogoliubov quasiparticles from bose-einstein condensate in a flat band system},\ }\href {https://doi.org/10.1103/PhysRevLett.131.226601} {\bibfield  {journal} {\bibinfo  {journal} {Phys. Rev. Lett.}\ }\textbf {\bibinfo {volume} {131}},\ \bibinfo {pages} {226601} (\bibinfo {year} {2023})}\BibitemShut {NoStop}%
\bibitem [{\citenamefont {Yin}\ \emph {et~al.}(2024)\citenamefont {Yin}, \citenamefont {Ye}, \citenamefont {He}, \citenamefont {Huang}, \citenamefont {Ke}, \citenamefont {Deng}, \citenamefont {Lu},\ and\ \citenamefont {Liu}}]{YIN20241660}%
  \BibitemOpen
  \bibfield  {author} {\bibinfo {author} {\bibfnamefont {S.}~\bibnamefont {Yin}}, \bibinfo {author} {\bibfnamefont {L.}~\bibnamefont {Ye}}, \bibinfo {author} {\bibfnamefont {H.}~\bibnamefont {He}}, \bibinfo {author} {\bibfnamefont {X.}~\bibnamefont {Huang}}, \bibinfo {author} {\bibfnamefont {M.}~\bibnamefont {Ke}}, \bibinfo {author} {\bibfnamefont {W.}~\bibnamefont {Deng}}, \bibinfo {author} {\bibfnamefont {J.}~\bibnamefont {Lu}},\ and\ \bibinfo {author} {\bibfnamefont {Z.}~\bibnamefont {Liu}},\ }\bibfield  {title} {\bibinfo {title} {Valley edge states as bound states in the continuum},\ }\href {https://doi.org/https://doi.org/10.1016/j.scib.2024.04.007} {\bibfield  {journal} {\bibinfo  {journal} {Science Bulletin}\ }\textbf {\bibinfo {volume} {69}},\ \bibinfo {pages} {1660} (\bibinfo {year} {2024})}\BibitemShut {NoStop}%
\bibitem [{\citenamefont {Wang}\ \emph {et~al.}(2022)\citenamefont {Wang}, \citenamefont {Wang},\ and\ \citenamefont {Ma}}]{BIC_NH}%
  \BibitemOpen
  \bibfield  {author} {\bibinfo {author} {\bibfnamefont {W.}~\bibnamefont {Wang}}, \bibinfo {author} {\bibfnamefont {X.}~\bibnamefont {Wang}},\ and\ \bibinfo {author} {\bibfnamefont {G.}~\bibnamefont {Ma}},\ }\bibfield  {title} {\bibinfo {title} {Extended state in a localized continuum},\ }\href {https://doi.org/10.1103/PhysRevLett.129.264301} {\bibfield  {journal} {\bibinfo  {journal} {Phys. Rev. Lett.}\ }\textbf {\bibinfo {volume} {129}},\ \bibinfo {pages} {264301} (\bibinfo {year} {2022})}\BibitemShut {NoStop}%
\bibitem [{\citenamefont {Qian}\ \emph {et~al.}(2024)\citenamefont {Qian}, \citenamefont {Zhang}, \citenamefont {Sun},\ and\ \citenamefont {Zhang}}]{Non-Abelian_BIC}%
  \BibitemOpen
  \bibfield  {author} {\bibinfo {author} {\bibfnamefont {L.}~\bibnamefont {Qian}}, \bibinfo {author} {\bibfnamefont {W.}~\bibnamefont {Zhang}}, \bibinfo {author} {\bibfnamefont {H.}~\bibnamefont {Sun}},\ and\ \bibinfo {author} {\bibfnamefont {X.}~\bibnamefont {Zhang}},\ }\bibfield  {title} {\bibinfo {title} {Non-abelian topological bound states in the continuum},\ }\href {https://doi.org/10.1103/PhysRevLett.132.046601} {\bibfield  {journal} {\bibinfo  {journal} {Phys. Rev. Lett.}\ }\textbf {\bibinfo {volume} {132}},\ \bibinfo {pages} {046601} (\bibinfo {year} {2024})}\BibitemShut {NoStop}%
\bibitem [{\citenamefont {Qiang}\ \emph {et~al.}()\citenamefont {Qiang}, \citenamefont {Ma},\ and\ \citenamefont {Song}}]{REv_QRW}%
  \BibitemOpen
  \bibfield  {author} {\bibinfo {author} {\bibfnamefont {X.}~\bibnamefont {Qiang}}, \bibinfo {author} {\bibfnamefont {S.}~\bibnamefont {Ma}},\ and\ \bibinfo {author} {\bibfnamefont {H.}~\bibnamefont {Song}},\ }\bibfield  {title} {\bibinfo {title} {Review on quantum walk computing: Theory, implementation, and application},\ }\href {https://doi.org/10.34133/icomputing.0097} {\bibfield  {journal} {\bibinfo  {journal} {Intelligent Computing}\ }\textbf {\bibinfo {volume} {0}}}\BibitemShut {NoStop}%
\bibitem [{\citenamefont {Aharonov}\ \emph {et~al.}(1993)\citenamefont {Aharonov}, \citenamefont {Davidovich},\ and\ \citenamefont {Zagury}}]{QRW_aharonov}%
  \BibitemOpen
  \bibfield  {author} {\bibinfo {author} {\bibfnamefont {Y.}~\bibnamefont {Aharonov}}, \bibinfo {author} {\bibfnamefont {L.}~\bibnamefont {Davidovich}},\ and\ \bibinfo {author} {\bibfnamefont {N.}~\bibnamefont {Zagury}},\ }\bibfield  {title} {\bibinfo {title} {Quantum random walks},\ }\href {https://doi.org/10.1103/PhysRevA.48.1687} {\bibfield  {journal} {\bibinfo  {journal} {Phys. Rev. A}\ }\textbf {\bibinfo {volume} {48}},\ \bibinfo {pages} {1687} (\bibinfo {year} {1993})}\BibitemShut {NoStop}%
\bibitem [{\citenamefont {Kempe}(2003)}]{QRW_Kempe}%
  \BibitemOpen
  \bibfield  {author} {\bibinfo {author} {\bibfnamefont {J.}~\bibnamefont {Kempe}},\ }\bibfield  {title} {\bibinfo {title} {Quantum random walks: An introductory overview},\ }\href {https://doi.org/10.1080/00107151031000110776} {\bibfield  {journal} {\bibinfo  {journal} {Contemporary Physics}\ }\textbf {\bibinfo {volume} {44}},\ \bibinfo {pages} {307} (\bibinfo {year} {2003})}\BibitemShut {NoStop}%
\bibitem [{\citenamefont {Nayak}\ and\ \citenamefont {Vishwanath}(2000)}]{nayak2000quantumwalkline}%
  \BibitemOpen
  \bibfield  {author} {\bibinfo {author} {\bibfnamefont {A.}~\bibnamefont {Nayak}}\ and\ \bibinfo {author} {\bibfnamefont {A.}~\bibnamefont {Vishwanath}},\ }\href {https://arxiv.org/abs/quant-ph/0010117} {\bibinfo {title} {Quantum walk on the line}} (\bibinfo {year} {2000}),\ \Eprint {https://arxiv.org/abs/quant-ph/0010117} {arXiv:quant-ph/0010117 [quant-ph]} \BibitemShut {NoStop}%
\bibitem [{\citenamefont {El~Sokhen}\ \emph {et~al.}(2024)\citenamefont {El~Sokhen}, \citenamefont {G\'omez-Le\'on}, \citenamefont {Adiyatullin}, \citenamefont {Randoux}, \citenamefont {Delplace},\ and\ \citenamefont {Amo}}]{Edge-anomalous}%
  \BibitemOpen
  \bibfield  {author} {\bibinfo {author} {\bibfnamefont {R.}~\bibnamefont {El~Sokhen}}, \bibinfo {author} {\bibfnamefont {A.}~\bibnamefont {G\'omez-Le\'on}}, \bibinfo {author} {\bibfnamefont {A.~F.}\ \bibnamefont {Adiyatullin}}, \bibinfo {author} {\bibfnamefont {S.}~\bibnamefont {Randoux}}, \bibinfo {author} {\bibfnamefont {P.}~\bibnamefont {Delplace}},\ and\ \bibinfo {author} {\bibfnamefont {A.}~\bibnamefont {Amo}},\ }\bibfield  {title} {\bibinfo {title} {Edge-dependent anomalous topology in synthetic photonic lattices subject to discrete step walks},\ }\href {https://doi.org/10.1103/PhysRevResearch.6.023282} {\bibfield  {journal} {\bibinfo  {journal} {Phys. Rev. Res.}\ }\textbf {\bibinfo {volume} {6}},\ \bibinfo {pages} {023282} (\bibinfo {year} {2024})}\BibitemShut {NoStop}%
\bibitem [{\citenamefont {Kitagawa}\ \emph {et~al.}(2010)\citenamefont {Kitagawa}, \citenamefont {Rudner}, \citenamefont {Berg},\ and\ \citenamefont {Demler}}]{Demler_2010}%
  \BibitemOpen
  \bibfield  {author} {\bibinfo {author} {\bibfnamefont {T.}~\bibnamefont {Kitagawa}}, \bibinfo {author} {\bibfnamefont {M.~S.}\ \bibnamefont {Rudner}}, \bibinfo {author} {\bibfnamefont {E.}~\bibnamefont {Berg}},\ and\ \bibinfo {author} {\bibfnamefont {E.}~\bibnamefont {Demler}},\ }\bibfield  {title} {\bibinfo {title} {Exploring topological phases with quantum walks},\ }\href {https://doi.org/10.1103/PhysRevA.82.033429} {\bibfield  {journal} {\bibinfo  {journal} {Phys. Rev. A}\ }\textbf {\bibinfo {volume} {82}},\ \bibinfo {pages} {033429} (\bibinfo {year} {2010})}\BibitemShut {NoStop}%
\bibitem [{\citenamefont {Rudner}\ and\ \citenamefont {Levitov}(2009)}]{Runder_PRL2009}%
  \BibitemOpen
  \bibfield  {author} {\bibinfo {author} {\bibfnamefont {M.~S.}\ \bibnamefont {Rudner}}\ and\ \bibinfo {author} {\bibfnamefont {L.~S.}\ \bibnamefont {Levitov}},\ }\bibfield  {title} {\bibinfo {title} {Topological transition in a non-hermitian quantum walk},\ }\href {https://doi.org/10.1103/PhysRevLett.102.065703} {\bibfield  {journal} {\bibinfo  {journal} {Phys. Rev. Lett.}\ }\textbf {\bibinfo {volume} {102}},\ \bibinfo {pages} {065703} (\bibinfo {year} {2009})}\BibitemShut {NoStop}%
\bibitem [{\citenamefont {Rudner}\ \emph {et~al.}(2013)\citenamefont {Rudner}, \citenamefont {Lindner}, \citenamefont {Berg},\ and\ \citenamefont {Levin}}]{Runder_PRX2013}%
  \BibitemOpen
  \bibfield  {author} {\bibinfo {author} {\bibfnamefont {M.~S.}\ \bibnamefont {Rudner}}, \bibinfo {author} {\bibfnamefont {N.~H.}\ \bibnamefont {Lindner}}, \bibinfo {author} {\bibfnamefont {E.}~\bibnamefont {Berg}},\ and\ \bibinfo {author} {\bibfnamefont {M.}~\bibnamefont {Levin}},\ }\bibfield  {title} {\bibinfo {title} {Anomalous edge states and the bulk-edge correspondence for periodically driven two-dimensional systems},\ }\href {https://doi.org/10.1103/PhysRevX.3.031005} {\bibfield  {journal} {\bibinfo  {journal} {Phys. Rev. X}\ }\textbf {\bibinfo {volume} {3}},\ \bibinfo {pages} {031005} (\bibinfo {year} {2013})}\BibitemShut {NoStop}%
\bibitem [{\citenamefont {Adiyatullin}\ \emph {et~al.}(2023)\citenamefont {Adiyatullin}, \citenamefont {Upreti}, \citenamefont {Lechevalier}, \citenamefont {Evain}, \citenamefont {Copie}, \citenamefont {Suret}, \citenamefont {Randoux}, \citenamefont {Delplace},\ and\ \citenamefont {Amo}}]{Topo_Floquet_PRL2023}%
  \BibitemOpen
  \bibfield  {author} {\bibinfo {author} {\bibfnamefont {A.~F.}\ \bibnamefont {Adiyatullin}}, \bibinfo {author} {\bibfnamefont {L.~K.}\ \bibnamefont {Upreti}}, \bibinfo {author} {\bibfnamefont {C.}~\bibnamefont {Lechevalier}}, \bibinfo {author} {\bibfnamefont {C.}~\bibnamefont {Evain}}, \bibinfo {author} {\bibfnamefont {F.}~\bibnamefont {Copie}}, \bibinfo {author} {\bibfnamefont {P.}~\bibnamefont {Suret}}, \bibinfo {author} {\bibfnamefont {S.}~\bibnamefont {Randoux}}, \bibinfo {author} {\bibfnamefont {P.}~\bibnamefont {Delplace}},\ and\ \bibinfo {author} {\bibfnamefont {A.}~\bibnamefont {Amo}},\ }\bibfield  {title} {\bibinfo {title} {Topological properties of floquet winding bands in a photonic lattice},\ }\href {https://doi.org/10.1103/PhysRevLett.130.056901} {\bibfield  {journal} {\bibinfo  {journal} {Phys. Rev. Lett.}\ }\textbf {\bibinfo {volume} {130}},\ \bibinfo {pages} {056901} (\bibinfo {year} {2023})}\BibitemShut {NoStop}%
\bibitem [{\citenamefont {Cedzich}\ \emph {et~al.}(2016)\citenamefont {Cedzich}, \citenamefont {Grünbaum}, \citenamefont {Stahl}, \citenamefont {Velázquez}, \citenamefont {Werner},\ and\ \citenamefont {Werner}}]{Cedzich_2016}%
  \BibitemOpen
  \bibfield  {author} {\bibinfo {author} {\bibfnamefont {C.}~\bibnamefont {Cedzich}}, \bibinfo {author} {\bibfnamefont {F.~A.}\ \bibnamefont {Grünbaum}}, \bibinfo {author} {\bibfnamefont {C.}~\bibnamefont {Stahl}}, \bibinfo {author} {\bibfnamefont {L.}~\bibnamefont {Velázquez}}, \bibinfo {author} {\bibfnamefont {A.~H.}\ \bibnamefont {Werner}},\ and\ \bibinfo {author} {\bibfnamefont {R.~F.}\ \bibnamefont {Werner}},\ }\bibfield  {title} {\bibinfo {title} {Bulk-edge correspondence of one-dimensional quantum walks},\ }\href {https://doi.org/10.1088/1751-8113/49/21/21LT01} {\bibfield  {journal} {\bibinfo  {journal} {Journal of Physics A: Mathematical and Theoretical}\ }\textbf {\bibinfo {volume} {49}},\ \bibinfo {pages} {21LT01} (\bibinfo {year} {2016})}\BibitemShut {NoStop}%
\bibitem [{\citenamefont {Asb\'oth}\ and\ \citenamefont {Obuse}(2013)}]{PhysRevB.88.121406}%
  \BibitemOpen
  \bibfield  {author} {\bibinfo {author} {\bibfnamefont {J.~K.}\ \bibnamefont {Asb\'oth}}\ and\ \bibinfo {author} {\bibfnamefont {H.}~\bibnamefont {Obuse}},\ }\bibfield  {title} {\bibinfo {title} {Bulk-boundary correspondence for chiral symmetric quantum walks},\ }\href {https://doi.org/10.1103/PhysRevB.88.121406} {\bibfield  {journal} {\bibinfo  {journal} {Phys. Rev. B}\ }\textbf {\bibinfo {volume} {88}},\ \bibinfo {pages} {121406} (\bibinfo {year} {2013})}\BibitemShut {NoStop}%
\bibitem [{\citenamefont {De~Nicola}\ \emph {et~al.}(2014)\citenamefont {De~Nicola}, \citenamefont {Sansoni}, \citenamefont {Crespi}, \citenamefont {Ramponi}, \citenamefont {Osellame}, \citenamefont {Giovannetti}, \citenamefont {Fazio}, \citenamefont {Mataloni},\ and\ \citenamefont {Sciarrino}}]{Quantum_simulation}%
  \BibitemOpen
  \bibfield  {author} {\bibinfo {author} {\bibfnamefont {F.}~\bibnamefont {De~Nicola}}, \bibinfo {author} {\bibfnamefont {L.}~\bibnamefont {Sansoni}}, \bibinfo {author} {\bibfnamefont {A.}~\bibnamefont {Crespi}}, \bibinfo {author} {\bibfnamefont {R.}~\bibnamefont {Ramponi}}, \bibinfo {author} {\bibfnamefont {R.}~\bibnamefont {Osellame}}, \bibinfo {author} {\bibfnamefont {V.}~\bibnamefont {Giovannetti}}, \bibinfo {author} {\bibfnamefont {R.}~\bibnamefont {Fazio}}, \bibinfo {author} {\bibfnamefont {P.}~\bibnamefont {Mataloni}},\ and\ \bibinfo {author} {\bibfnamefont {F.}~\bibnamefont {Sciarrino}},\ }\bibfield  {title} {\bibinfo {title} {Quantum simulation of bosonic-fermionic noninteracting particles in disordered systems via a quantum walk},\ }\href {https://doi.org/10.1103/PhysRevA.89.032322} {\bibfield  {journal} {\bibinfo  {journal} {Phys. Rev. A}\ }\textbf {\bibinfo {volume} {89}},\ \bibinfo {pages} {032322} (\bibinfo {year} {2014})}\BibitemShut {NoStop}%
\bibitem [{\citenamefont {Schreiber}\ \emph {et~al.}(2010)\citenamefont {Schreiber}, \citenamefont {Cassemiro}, \citenamefont {Poto\ifmmode~\check{c}\else \v{c}\fi{}ek}, \citenamefont {G\'abris}, \citenamefont {Mosley}, \citenamefont {Andersson}, \citenamefont {Jex},\ and\ \citenamefont {Silberhorn}}]{QRW_Coin}%
  \BibitemOpen
  \bibfield  {author} {\bibinfo {author} {\bibfnamefont {A.}~\bibnamefont {Schreiber}}, \bibinfo {author} {\bibfnamefont {K.~N.}\ \bibnamefont {Cassemiro}}, \bibinfo {author} {\bibfnamefont {V.}~\bibnamefont {Poto\ifmmode~\check{c}\else \v{c}\fi{}ek}}, \bibinfo {author} {\bibfnamefont {A.}~\bibnamefont {G\'abris}}, \bibinfo {author} {\bibfnamefont {P.~J.}\ \bibnamefont {Mosley}}, \bibinfo {author} {\bibfnamefont {E.}~\bibnamefont {Andersson}}, \bibinfo {author} {\bibfnamefont {I.}~\bibnamefont {Jex}},\ and\ \bibinfo {author} {\bibfnamefont {C.}~\bibnamefont {Silberhorn}},\ }\bibfield  {title} {\bibinfo {title} {Photons walking the line: A quantum walk with adjustable coin operations},\ }\href {https://doi.org/10.1103/PhysRevLett.104.050502} {\bibfield  {journal} {\bibinfo  {journal} {Phys. Rev. Lett.}\ }\textbf {\bibinfo {volume} {104}},\ \bibinfo {pages} {050502} (\bibinfo {year} {2010})}\BibitemShut {NoStop}%
\bibitem [{\citenamefont {Childs}\ \emph {et~al.}(2013)\citenamefont {Childs}, \citenamefont {Gosset},\ and\ \citenamefont {Webb}}]{science_computation}%
  \BibitemOpen
  \bibfield  {author} {\bibinfo {author} {\bibfnamefont {A.~M.}\ \bibnamefont {Childs}}, \bibinfo {author} {\bibfnamefont {D.}~\bibnamefont {Gosset}},\ and\ \bibinfo {author} {\bibfnamefont {Z.}~\bibnamefont {Webb}},\ }\bibfield  {title} {\bibinfo {title} {Universal computation by multiparticle quantum walk},\ }\href {https://doi.org/10.1126/science.1229957} {\bibfield  {journal} {\bibinfo  {journal} {Science}\ }\textbf {\bibinfo {volume} {339}},\ \bibinfo {pages} {791} (\bibinfo {year} {2013})}\BibitemShut {NoStop}%
\bibitem [{\citenamefont {Lovett}\ \emph {et~al.}(2010)\citenamefont {Lovett}, \citenamefont {Cooper}, \citenamefont {Everitt}, \citenamefont {Trevers},\ and\ \citenamefont {Kendon}}]{Computation_QRW1}%
  \BibitemOpen
  \bibfield  {author} {\bibinfo {author} {\bibfnamefont {N.~B.}\ \bibnamefont {Lovett}}, \bibinfo {author} {\bibfnamefont {S.}~\bibnamefont {Cooper}}, \bibinfo {author} {\bibfnamefont {M.}~\bibnamefont {Everitt}}, \bibinfo {author} {\bibfnamefont {M.}~\bibnamefont {Trevers}},\ and\ \bibinfo {author} {\bibfnamefont {V.}~\bibnamefont {Kendon}},\ }\bibfield  {title} {\bibinfo {title} {Universal quantum computation using the discrete-time quantum walk},\ }\href {https://doi.org/10.1103/PhysRevA.81.042330} {\bibfield  {journal} {\bibinfo  {journal} {Phys. Rev. A}\ }\textbf {\bibinfo {volume} {81}},\ \bibinfo {pages} {042330} (\bibinfo {year} {2010})}\BibitemShut {NoStop}%
\bibitem [{\citenamefont {Colandrea}\ \emph {et~al.}(2023)\citenamefont {Colandrea}, \citenamefont {Babazadeh}, \citenamefont {Dauphin}, \citenamefont {Massignan}, \citenamefont {Marrucci},\ and\ \citenamefont {Cardano}}]{Entag_QRW_Ex_2023}%
  \BibitemOpen
  \bibfield  {author} {\bibinfo {author} {\bibfnamefont {F.~D.}\ \bibnamefont {Colandrea}}, \bibinfo {author} {\bibfnamefont {A.}~\bibnamefont {Babazadeh}}, \bibinfo {author} {\bibfnamefont {A.}~\bibnamefont {Dauphin}}, \bibinfo {author} {\bibfnamefont {P.}~\bibnamefont {Massignan}}, \bibinfo {author} {\bibfnamefont {L.}~\bibnamefont {Marrucci}},\ and\ \bibinfo {author} {\bibfnamefont {F.}~\bibnamefont {Cardano}},\ }\bibfield  {title} {\bibinfo {title} {Ultra-long quantum walks via spin--orbit photonics},\ }\href {https://doi.org/10.1364/OPTICA.474542} {\bibfield  {journal} {\bibinfo  {journal} {Optica}\ }\textbf {\bibinfo {volume} {10}},\ \bibinfo {pages} {324} (\bibinfo {year} {2023})}\BibitemShut {NoStop}%
\bibitem [{\citenamefont {Xia}\ \emph {et~al.}(2020)\citenamefont {Xia}, \citenamefont {Liu}, \citenamefont {Nie}, \citenamefont {Fu}, \citenamefont {Wan},\ and\ \citenamefont {Kong}}]{QRW_algo1}%
  \BibitemOpen
  \bibfield  {author} {\bibinfo {author} {\bibfnamefont {F.}~\bibnamefont {Xia}}, \bibinfo {author} {\bibfnamefont {J.}~\bibnamefont {Liu}}, \bibinfo {author} {\bibfnamefont {H.}~\bibnamefont {Nie}}, \bibinfo {author} {\bibfnamefont {Y.}~\bibnamefont {Fu}}, \bibinfo {author} {\bibfnamefont {L.}~\bibnamefont {Wan}},\ and\ \bibinfo {author} {\bibfnamefont {X.}~\bibnamefont {Kong}},\ }\bibfield  {title} {\bibinfo {title} {Random walks: A review of algorithms and applications},\ }\href {https://doi.org/10.1109/TETCI.2019.2952908} {\bibfield  {journal} {\bibinfo  {journal} {IEEE Trans. Emerg. Top. Comput. Intell.}\ }\textbf {\bibinfo {volume} {4}},\ \bibinfo {pages} {95} (\bibinfo {year} {2020})}\BibitemShut {NoStop}%
\bibitem [{\citenamefont {Shenvi}\ \emph {et~al.}(2003)\citenamefont {Shenvi}, \citenamefont {Kempe},\ and\ \citenamefont {Whaley}}]{QRW_algo}%
  \BibitemOpen
  \bibfield  {author} {\bibinfo {author} {\bibfnamefont {N.}~\bibnamefont {Shenvi}}, \bibinfo {author} {\bibfnamefont {J.}~\bibnamefont {Kempe}},\ and\ \bibinfo {author} {\bibfnamefont {K.~B.}\ \bibnamefont {Whaley}},\ }\bibfield  {title} {\bibinfo {title} {Quantum random-walk search algorithm},\ }\href {https://doi.org/10.1103/PhysRevA.67.052307} {\bibfield  {journal} {\bibinfo  {journal} {Phys. Rev. A}\ }\textbf {\bibinfo {volume} {67}},\ \bibinfo {pages} {052307} (\bibinfo {year} {2003})}\BibitemShut {NoStop}%
\bibitem [{\citenamefont {Schmitz}\ \emph {et~al.}(2009)\citenamefont {Schmitz}, \citenamefont {Matjeschk}, \citenamefont {Schneider}, \citenamefont {Glueckert}, \citenamefont {Enderlein}, \citenamefont {Huber},\ and\ \citenamefont {Schaetz}}]{trap_ion_EXP}%
  \BibitemOpen
  \bibfield  {author} {\bibinfo {author} {\bibfnamefont {H.}~\bibnamefont {Schmitz}}, \bibinfo {author} {\bibfnamefont {R.}~\bibnamefont {Matjeschk}}, \bibinfo {author} {\bibfnamefont {C.}~\bibnamefont {Schneider}}, \bibinfo {author} {\bibfnamefont {J.}~\bibnamefont {Glueckert}}, \bibinfo {author} {\bibfnamefont {M.}~\bibnamefont {Enderlein}}, \bibinfo {author} {\bibfnamefont {T.}~\bibnamefont {Huber}},\ and\ \bibinfo {author} {\bibfnamefont {T.}~\bibnamefont {Schaetz}},\ }\bibfield  {title} {\bibinfo {title} {Quantum walk of a trapped ion in phase space},\ }\href {https://doi.org/10.1103/PhysRevLett.103.090504} {\bibfield  {journal} {\bibinfo  {journal} {Phys. Rev. Lett.}\ }\textbf {\bibinfo {volume} {103}},\ \bibinfo {pages} {090504} (\bibinfo {year} {2009})}\BibitemShut {NoStop}%
\bibitem [{\citenamefont {Blatt}\ and\ \citenamefont {Roos}(2012)}]{Blatt2012}%
  \BibitemOpen
  \bibfield  {author} {\bibinfo {author} {\bibfnamefont {R.}~\bibnamefont {Blatt}}\ and\ \bibinfo {author} {\bibfnamefont {C.~F.}\ \bibnamefont {Roos}},\ }\bibfield  {title} {\bibinfo {title} {Quantum simulations with trapped ions},\ }\href {https://doi.org/10.1038/nphys2252} {\bibfield  {journal} {\bibinfo  {journal} {Nature Physics}\ }\textbf {\bibinfo {volume} {8}},\ \bibinfo {pages} {277} (\bibinfo {year} {2012})}\BibitemShut {NoStop}%
\bibitem [{\citenamefont {Karski}\ \emph {et~al.}(2009)\citenamefont {Karski}, \citenamefont {Förster}, \citenamefont {Choi}, \citenamefont {Steffen}, \citenamefont {Alt}, \citenamefont {Meschede},\ and\ \citenamefont {Widera}}]{optical_lattice_QRW}%
  \BibitemOpen
  \bibfield  {author} {\bibinfo {author} {\bibfnamefont {M.}~\bibnamefont {Karski}}, \bibinfo {author} {\bibfnamefont {L.}~\bibnamefont {Förster}}, \bibinfo {author} {\bibfnamefont {J.-M.}\ \bibnamefont {Choi}}, \bibinfo {author} {\bibfnamefont {A.}~\bibnamefont {Steffen}}, \bibinfo {author} {\bibfnamefont {W.}~\bibnamefont {Alt}}, \bibinfo {author} {\bibfnamefont {D.}~\bibnamefont {Meschede}},\ and\ \bibinfo {author} {\bibfnamefont {A.}~\bibnamefont {Widera}},\ }\bibfield  {title} {\bibinfo {title} {Quantum walk in position space with single optically trapped atoms},\ }\href {https://doi.org/10.1126/science.1174436} {\bibfield  {journal} {\bibinfo  {journal} {Science}\ }\textbf {\bibinfo {volume} {325}},\ \bibinfo {pages} {174} (\bibinfo {year} {2009})}\BibitemShut {NoStop}%
\bibitem [{\citenamefont {Tude}\ and\ \citenamefont {de~Oliveira}(2022)}]{Tude_2022}%
  \BibitemOpen
  \bibfield  {author} {\bibinfo {author} {\bibfnamefont {L.~T.}\ \bibnamefont {Tude}}\ and\ \bibinfo {author} {\bibfnamefont {M.~C.}\ \bibnamefont {de~Oliveira}},\ }\bibfield  {title} {\bibinfo {title} {Temperature and entanglement of the three-state quantum walk},\ }\href {https://doi.org/10.1088/2058-9565/ac6a05} {\bibfield  {journal} {\bibinfo  {journal} {Quantum Science and Technology}\ }\textbf {\bibinfo {volume} {7}},\ \bibinfo {pages} {035009} (\bibinfo {year} {2022})}\BibitemShut {NoStop}%
\bibitem [{\citenamefont {Inui}\ \emph {et~al.}(2005)\citenamefont {Inui}, \citenamefont {Konno},\ and\ \citenamefont {Segawa}}]{three-state}%
  \BibitemOpen
  \bibfield  {author} {\bibinfo {author} {\bibfnamefont {N.}~\bibnamefont {Inui}}, \bibinfo {author} {\bibfnamefont {N.}~\bibnamefont {Konno}},\ and\ \bibinfo {author} {\bibfnamefont {E.}~\bibnamefont {Segawa}},\ }\bibfield  {title} {\bibinfo {title} {One-dimensional three-state quantum walk},\ }\href {https://doi.org/10.1103/PhysRevE.72.056112} {\bibfield  {journal} {\bibinfo  {journal} {Phys. Rev. E}\ }\textbf {\bibinfo {volume} {72}},\ \bibinfo {pages} {056112} (\bibinfo {year} {2005})}\BibitemShut {NoStop}%
\bibitem [{\citenamefont {Georgi}(2000)}]{book-lie}%
  \BibitemOpen
  \bibfield  {author} {\bibinfo {author} {\bibfnamefont {H.}~\bibnamefont {Georgi}},\ }\href@noop {} {\emph {\bibinfo {title} {Lie algebras in particle physics: from isospin to unified theories}}}\ (\bibinfo  {publisher} {Taylor \& Francis},\ \bibinfo {year} {2000})\BibitemShut {NoStop}%
\bibitem [{\citenamefont {Gell-Mann}(1962)}]{Gellman}%
  \BibitemOpen
  \bibfield  {author} {\bibinfo {author} {\bibfnamefont {M.}~\bibnamefont {Gell-Mann}},\ }\bibfield  {title} {\bibinfo {title} {Symmetries of baryons and mesons},\ }\href {https://doi.org/10.1103/PhysRev.125.1067} {\bibfield  {journal} {\bibinfo  {journal} {Phys. Rev.}\ }\textbf {\bibinfo {volume} {125}},\ \bibinfo {pages} {1067} (\bibinfo {year} {1962})}\BibitemShut {NoStop}%
\bibitem [{\citenamefont {Kaiser}(2022)}]{SU3}%
  \BibitemOpen
  \bibfield  {author} {\bibinfo {author} {\bibfnamefont {N.}~\bibnamefont {Kaiser}},\ }\bibfield  {title} {\bibinfo {title} {Solving the matrix exponential function for the lie groups su(3), su(4) and sp(2)},\ }\href {https://doi.org/10.1140/epja/s10050-022-00816-5} {\bibfield  {journal} {\bibinfo  {journal} {The European Physical Journal A}\ }\textbf {\bibinfo {volume} {58}},\ \bibinfo {pages} {170} (\bibinfo {year} {2022})}\BibitemShut {NoStop}%
\bibitem [{\citenamefont {Fukui}\ \emph {et~al.}(2005)\citenamefont {Fukui}, \citenamefont {Hatsugai},\ and\ \citenamefont {Suzuki}}]{Fuki}%
  \BibitemOpen
  \bibfield  {author} {\bibinfo {author} {\bibfnamefont {T.}~\bibnamefont {Fukui}}, \bibinfo {author} {\bibfnamefont {Y.}~\bibnamefont {Hatsugai}},\ and\ \bibinfo {author} {\bibfnamefont {H.}~\bibnamefont {Suzuki}},\ }\bibfield  {title} {\bibinfo {title} {Chern numbers in discretized brillouin zone: Efficient method of computing (spin) hall conductances},\ }\href {https://doi.org/10.1143/JPSJ.74.1674} {\bibfield  {journal} {\bibinfo  {journal} {Journal of the Physical Society of Japan}\ }\textbf {\bibinfo {volume} {74}},\ \bibinfo {pages} {1674} (\bibinfo {year} {2005})}\BibitemShut {NoStop}%
\bibitem [{See Supplemental Material at [url] for more details()}]{supp}%
  \BibitemOpen
  See Supplemental Material at [url] for more details,\ \href@noop {} {}\BibitemShut {NoStop}%
\end{thebibliography}%



\end{document}